\documentclass[11pt,english,a4paper] {article}
\usepackage{latexsym,epsfig,amsbsy}

\def \del {\partial}
\def \Rarr {\Rightarrow}
\def \12{\frac{1}{2}}
\def \air {\medskip \noindent}
\newcommand{\beq}{\begin{equation}}
\newcommand{\eeq}{\end{equation}}

\title{
\begin{flushright}
\small
USITP-00-09\\
hep-th/0007101
\normalsize
\end{flushright}
\bf{A pedestrian approach to\\
 high energy limits of branes and other
gravitational systems}} 
\author{
	Ulf Lindstr\"om\footnote{ul@physto.se}
	\\ \emph{Institute of Theoretical Physics, University of Stockholm}
	\and
	Harald G. Svendsen\footnote{h.g.svendsen@fys.uio.no}
	\\ \emph{Department of Physics, University of Oslo}
}

\begin{document}

\maketitle

\abstract
In this article we study limits of models that contain a dimensionful
parameter such as the mass of the relativistic point-particle.
The limits are analogous to the massless limit of the particle 
and may be thought of as high energy limits. We present the ideas 
and work through several examples in a (hopefully) pedagogical
manner.  Along the way we derive several new results.


\section{Introduction}

The purpose of this article is to give a description and demonstration of 
 methods for deriving high energy limits of various action
integrals. It is meant to be readable for a broad range of
theoretical physicists, and no expert background is required.

The  models we study are described by an action
integral of the form $S = T \int dx {\cal L}(\phi,\del\phi)$, $T$ being some 
constant and $L=T{\cal L}$ 
the  Lagrangian density.\footnote{Throughout 
this article the Lagrangian is written as 
$L = T{\cal L}$, where $T$ is the constant relevant for the theory
under study.}
The constant $T$
can typically be neglected at high energy scales, and hence the $T\to 0$
limit may be viewed as a high energy limit. It is this
limit which is the focus of this article.

Apart from giving an approximate description of the high energy
physics, this limit  is also interesting for other reasons.
First, it may be that the limit  represents a physical situation
regardless of the energy scale, in the same sense that photons can
be seen as massless particles. Second, the limits typically lead to
conformally invariant  theories, and one may thus hope to be
able to study the original models as broken conformal ones, a point
of view which has often proven fruitful.

\subsection{Constrained systems}
\label{sec:constrained_systems}

Since one of the methods extensively makes use of the theory for constrained 
systems, which dates back to Dirac \cite{dirac:1964}, it is useful
to briefly review this theory before entering the main subject of
this article. A more detailed description is found in e.g.
\cite{dirac:1964,henneaux:1992,marnelius:1982}.

Suppose you have a Lagrangian density $L$ which depends on
some fields
$\phi^i(x)$ and their derivatives,  i.e. $L=L(\phi^i,
\del_a\phi^i)$, (the index $i$ is not (necessarily) a component
index, but used only to keep track of the different fields).
Suppose furthermore that you have calculated the canonical
conjugate momenta to the fields $\phi^i(x)$,
\beq
  \label{eq:canonical_momentum}
  \pi^i = \frac{\del L}{\del \dot{\phi}^i}.
\eeq
Then the \emph{naive} Hamiltonian is given as usual by
\beq
  H_{naive} 
	= H_{naive}(\phi, \nabla\phi) 
	= \pi_i\dot{\phi}^i - L(\phi, \del\phi),
\eeq
where $\nabla$ signifies $\emph{spatial}$ derivatives, and $\del$
signifies  all partial derivatives. 

In cases where (\ref{eq:canonical_momentum}) can be inverted to give 
$\dot{\phi}^i = \dot{\phi}^i(\phi^i, \nabla\phi^i,\pi^i)$,
the naive Hamiltonian can easily be derived just by substitution of 
$\dot{\phi}^i$
as a function of $\pi^i$, $\phi^i$ and $\nabla\phi^i$.

On the other hand, if (\ref{eq:canonical_momentum}) is \emph{not}
invertible, we have a \emph{constrained system}.
The non-invertibility means that the  
momenta $\pi^i$
will not be independent, and hence that there exist some relations 
between them. These relations can be expressed as relations
$\theta^I_k(\phi, \pi, \nabla \phi)=0$, $k=1,\dots ,M$, where $M$ is 
the number of independent functions $\theta$. These functions are
called the
\emph{primary constraints} (hence the label $I$,) since they follow
directly from the definition of the momenta. 

Suppose that $i$ can take $N$ different values (i.e. there
are $N$ field variables). Then define the $N\times N$ matrix 
$C_{ij}\equiv \frac{\del^2 L}{\del\dot{\phi}^i\del\dot{\phi}^j}$. If
$R$ is the \emph{rank} of this matrix, then the number of independent
primary constraints can be shown to be  $N-R$.

For constrained systems the naive Hamiltonian is not
unique, since any linear combination of the constraint
functions $\theta^I$ can be added. This fact has implications for
Hamilton's equations, in which the naive Hamiltonian has to be
substituted by the \emph{extended Hamiltonian}
\beq
  H_I \equiv H_{naive} + \lambda_I^k\theta^I_k,
\eeq 
where $\lambda_I^k$ are
coefficients that do not depend on $\phi$ and $\pi$. (No summation
over $I$ implied.) These are
\emph{Lagrange multipliers}.

\paragraph{Poisson brackets}

Consider two functions $F$ and $G$ that are constructed
from $\phi^i$ and the conjugate momenta $\pi^i$. 
Writing $F$ as short for $F(\phi(x), \pi(x))$, 
the Poisson bracket of $F$ and $G$ can be defined by means of  
functional derivatives as
\beq
  \label{eq:poisson_bracket}
  \left\{F, G\right\}
   \equiv 
	\frac{\delta F}{ \delta \phi^i}
	\frac{\delta G}{\delta \pi_i} -
	\frac{\delta F}{ \delta \pi_i}
	\frac{\delta G}{\delta \phi^i}.
\eeq

When working with Poisson brackets in the present circumstance the
following is important: 
\emph{Poisson brackets must be evaluated before making use of the
constraint equations}. In other words, calculations should be
performed in  phase space, and with restriction to the constraint
surface ($\theta=0$) only at the very end.
To emphasize this point Dirac
introduced a notation  with a \emph{weak equality} sign
``$\approx$'', and wrote the constraint equations as
\beq
  \theta^I_m(\phi,\pi) \approx 0.
\eeq
This makes a difference, for even though $\theta(\phi,\pi)$ is
dynamically zero (i.e. zero when $\phi$ and $\pi$ satisfy 
Hamilton's equations) it is not zero throughout phase space.

\air
The requirement that the primary constraints should hold at all
times, i.e. time derivatives also being weakly zero, 
leads to a  set of consistency conditions
$\{\theta^I_m, H_I\}\approx 0$, or more explicitly,
\beq
  \label{eq:consistency-conditions}
  \{\theta^I_m, H_{naive}\} + \lambda_I^n\{\theta^I_m, \theta^I_n\}
	\approx 0.
\eeq
This may lead to a set of \emph{secondary constraints},
$\theta^{II}_m\approx 0$, and subsequently, a new set of 
consistency equations. 

When the full constraint structure has been found (i.e. when the
consistency conditions do not give anything new), the \emph{total
Hamiltonian} can be written
\beq
  H = H_{naive} + \lambda^i \theta_i,
\eeq
where $\lambda^i$ is a set of independent Lagrange multipliers, and
$\theta_i$ is the full set of constraints.

\subsection{The Naive Hamiltonian for diff invariant models}
\label{sec:diff_hamilton}

In this section it is demonstrated that theories which are invariant
under diffeomorphisms have a naive Hamiltonian that
is dynamically zero, i.e. zero when the field equations are imposed.

A \emph{diffeormorphism} is a general coordinate transformation of
the form
\[
  x \to \tilde{x} = \tilde{x}(x),
\]
where $\tilde{x}(x)$ is a differentiable function of $x$. For
models such as the point-particle whose world-line is embedded in
spacetime, a world-volume diffeomorphism is often called a
reparametrization.

\air
Consider the infinitesimal diffemorphism,
\beq
  x^a \to \tilde{x}^a = x^a + \xi^a,
\eeq
which gives the Jacobi determinant
\beq
  \label{eq:jacobideterminant}
  J \equiv \det( \frac{\del x^b}{\del\tilde{x}^a} )
	= \det(\delta^b_a - \del_a \xi^b)
	= 1 - \del_a\xi^a +....
\eeq
Under this change of variables the action integral transforms into
\beq
  S\to \tilde{S} = \int d\tilde{x} L(\tilde{\phi}(\tilde{x}), 
	\tilde{\del} \tilde{\phi}(\tilde{x}))
	= \int dx |J^{-1}| L(\tilde{\phi}(\tilde{x}),
	\tilde{\del}\tilde{\phi}(\tilde{x})).
\eeq

Let the transformation of the fields be written in the somewhat unusual form
\beq
  \phi^i(x) \to \tilde{\phi}^i(\tilde{x}) = \phi^i(x) +
	\epsilon^i(x).
\eeq
This defines $\epsilon^i$.  For scalars this notation
will give  $\epsilon=0$.

The derivatives of the fields transform as
\begin{eqnarray}
  \nonumber
  \tilde{\del}_b\tilde{\phi^i}(\tilde{x}) &=& 
	\frac{\del x^c}{\del\tilde{x}^b} \del_c (\phi^i(x) + \epsilon^i)
  \\
  &\cong& \del_b\phi^i(x) + \del_b\epsilon^i(x) - \del_b \xi^c\del_c\phi(x).
\end{eqnarray}
The transformed action integral may now be Taylor expanded and
rewritten in the following way:
\begin{eqnarray}
  \nonumber
  \tilde{S} 
  &=& \int dx (1+\del_c \xi^c)\left(
	L(\phi,\del\phi^i) + \frac{\del L}{\del\phi^i}\epsilon^i
	+\frac{\del L}{\del(\del_a\phi^i)}
	(\del_a\epsilon^i-\del_a \xi^b\del_b\phi^i) \right)
  \\ \nonumber
  &=& \int dx \bigg( L - \underbrace{ \left[
	\del_a(\frac{\del L}{\del(\del_a\phi^i)}) 
	- \frac{\del L}{\del\phi^i}\right]}_{=\psi_i} \epsilon^i
	+ \del_a \xi^b \underbrace{ \left[
	\delta^a_b L - \frac{\del L}{\del (\del_a\phi^i)}\del_b\phi^i
	\right]}_{\equiv T^a_{~b}} \bigg)
  \\ \label{eq:diff-varaction}
  &=& S + \int dx \left(\del_a \xi^b T^a_{~b}- \psi_i\epsilon^i\right),
\end{eqnarray}
where a partial integration has been performed with the assumption
that the fields vanish at infinity.

$T^a_{~b}$ is recognized as the \emph{canonical energy-momentum
tensor}. Of special interest here is the fact that
$T^0_{~0}= -H_{naive}$, by definition. 
Also, $\psi_i = 0$ is recognized as the \emph{Euler-Lagrange equations}.

For diff invariant theories $\delta S = \tilde{S}-S = 0$, which
gives the condition 
\beq
  \int dx \left[ T^a_{~b}\del_a \xi^b - \psi_i\epsilon^i\right] = 0. 
\eeq

Consider models made up by tensor fields of zero, first and second
rank, i.e. let  
$\phi^i = \{\phi,A_a,F_{ab}\}$\footnote{In general, each type of
field carries an index denumbering the fields of this type. For
ease of notation this index is suppressed.}, and define
$\Lambda^b_a\equiv \frac{\del x^b}{\del \tilde{x}^a} 
=\delta^b_a-\del_a \xi^b$.

For the scalar $\phi$, $\epsilon^\phi=0$.
For the vector field, 
$\tilde{A}_a(\tilde{x})= \Lambda^b_a A_b(x) 
= (\delta^b_a-\del_a \xi^b) A_b(x)$,
which gives
$\epsilon^A_b = -\del_b \xi^a A_a$.
For second rank tensors
$\tilde{F}_{ab}(\tilde{x}) = \Lambda^c_a\Lambda^d_b F_{cd}(x)$,
which gives
$\epsilon^F_{ab} = -\del_c \xi^d (\delta^c_a F_{db} + \delta^c_b
F_{ad})$.
These contributions add when all fields are present in the
Lagrangian.

For actions that depend on scalars, vectors and second rank tensors,
the diff symmetry criterion may thus be written
\beq
  \int dx \del_a \xi^b \left[ T^a_{~b}
	+ \psi^{a} A_b
	+ \psi^{ac}F_{bc}+\psi^{da}F_{db}
	\right]=0,
\eeq
where $\psi^a$ are the Euler-Lagrange equations associated with $A_a$
and $\psi^{ab}$ are the Euler-Lagrange equations associated with
$F_{ab}$. This equation should be true for arbitrary $\xi^a$, so
\beq
  T^a_{~b} = 
	- \psi^{a} A_b
	- \psi^{ac}F_{bc}-\psi^{da}F_{db},
\eeq
and specially
\beq
  \label{eq:diff_invariant_hamiltonian}
  H_{naive} = -T^0_{~0}
  = \psi^{0} A_0 + \psi^{0c}F_{0c}+\psi^{d0}F_{d0}
\eeq
As is well known, the energy-momentum tensor may be modified by
terms proportional to the field equations. (It is the Noether
current for translations and only conserved on-shell, in general).
Hence, with the appropriately redefined energy momentum tensor the
corresponding Hamiltonian vanishes. In the terminology
introduced in Section (1.1) we may rephrase this as saying that the
Hamiltonian is
dynamically zero. 

The above discussion quantifies a well known result which is only
known to us as ``folk lore'', except when there are only scalars
present, for which case it was proven by von Unge \cite{unge:1994}.
The above is an adaption of his proof. Our result is not completely 
general, since only tensor fields have been considered and  not, 
e.g., spinors.

\section{Two methods}

The starting point is an action of the form
\beq
  \label{eq:generalaction}
  S = T\int dx {\cal L}(\phi, \partial \phi),
\eeq
where $T$ is some constant with dimensions, like mass or string tension.
(As mentioned in the introduction, it is a basic assumption that the
Lagrangian can be written on this form,
$L=T{\cal L}$.)
The action (\ref{eq:generalaction}) is clearly not suitable for
studying the $T \to 0$ limit and  
the philosophy is to search for an action that is classically
equivalent to (\ref{eq:generalaction}) as long as $T\neq 0$, but also well 
defined for $T=0$. This new action (with $T=0$ inserted) may then
be  treated as a $T\to 0$ limit of the original model.
The two methods presented below are systematic ways for finding such
actions. The ideas have been around for quite
some time, but we have only been able to find what
we believe is the original reference for the second one. (Ref
\cite{marnelius:1982})

\subsection{Auxiliary field}
\label{sec:method-I}

This is the simplest approach, and involves the introduction of an 
auxiliary field $\chi$. 
A reference for this method is \cite{karlhede:1986}. The idea
is to define a new action integral,
\beq
  \label{eq:chiaction}
  S_\chi = \frac{1}{2} \int dx (\chi {\cal L}^2 + \frac{T^2}{\chi}).
\eeq
This action is equivalent to (\ref{eq:generalaction}), which can be shown
explicitly by solving the equations of motion for $\chi$:
\begin{eqnarray}
  \nonumber
  \delta \chi \Rarr \quad \delta S_\chi & = 
  & \frac{1}{2}\int dx (\delta\chi {\cal L}^2 - \frac{T^2}{\chi^2}\delta\chi)
  \\ \nonumber
  & = & \frac{1}{2} \int dx ({\cal L}^2 - \frac{T^2}{\chi^2})\delta\chi.
\end{eqnarray}
Using Hamilton's principle and demanding
$\delta S_\chi = 0$ for arbitrary variations $\delta \chi$ gives
\begin{eqnarray}
  \nonumber
  {\cal L}^2 - \frac{T^2}{\chi^2} & = & 0
  \\
  \chi & = & \frac{T}{{\cal L}}; \qquad \mbox{when}\quad T \neq 0.
\end{eqnarray}
This expression for $\chi$ substituted back into
(\ref{eq:chiaction}) gives
\[
  S_\chi = \frac{1}{2}\int dx(\frac{T}{{\cal L}}{\cal L}^2 
  + \frac{{\cal L}}{T}T^2) = T\int dx {\cal L} = S.
\]
Thus the two actions $S$ and $S_\chi$ are equivalent for $T \neq 0$. In
addition $S_\chi$ allows for the $T \to 0$ limit simply
by setting $T=0$ in the action. Doing so gives
\beq
  \label{eq:chi0action}
  S_\chi^{T=0} = \frac{1}{2} \int dx \chi {\cal L}^2.
\eeq
One natural question is now what this new field $\chi$ really
represents.  From the current point of view it is impossible to
say more than has already been said -- that it is useful for the
calculations. Hence the name \emph{auxiliary} field.

Note however, that in the simplest case of a massless particle 
$\chi$ should be interpreted as the einbein, as will become
apparent below.

A general remark on symmetry properties can be made already here. 
Consider diffeomorphism invariance. 
The integral measure transforms as 
$dx\to dxJ^{-1}$, where $J$ is the Jacobi determinant
as defined by equation (\ref{eq:jacobideterminant}). 
If the original action is to be diff invariant, the
Lagrangian must transform as a density, i.e. 
${\cal L} \to J {\cal L}$. 
$S_\chi$ is then also diff invariant if
$\chi$  transforms as an inverse density 
(i.e. scalar density of weight $-1$). 
Since $\chi$ was introduced as an
auxiliary field with no \emph{a priori} physical interpretation, this
transformation property is something that can be imposed on $\chi$.

\subsubsection{Dynamics}

The variation of $\chi$ gives one equation of motion,
\beq
  \label{eq:chi-motion}
  \delta\chi \Rarr\qquad {\cal L}^2=0 \quad\Rarr\quad {\cal L}=0.
\eeq
The result of a variation in $\phi^i$, on the other hand, depends on
the form of ${\cal L}$:
\begin{eqnarray}
  \nonumber
  \delta\phi^i \Rarr\qquad \delta S &=&
	\12\int dx \chi 2{\cal L}\delta{\cal L}
  \\ &=& \nonumber
  \int dx \chi {\cal L}\left[ \frac{\del {\cal
	L}}{\del\phi^i}\delta\phi^i
	+\frac{\del{\cal
	L}}{\del(\del_a\phi^i)}\del_a\delta\phi^i\right]
  \\ &=& \nonumber
  \int dx \left[\chi{\cal L}\frac{\del{\cal L}}{\del\phi^i} -
  \del_a\left(\chi{\cal L}\frac{\del{\cal
  L}}{\del(\del_a\phi^i)}\right)\right]\delta\phi^i.
\end{eqnarray}
The field equation is found by demanding $\delta S=0$ for arbitrary
$\delta\phi^i$. The result is
\beq
  \label{eq:phi-motion}
  \delta\phi \Rarr\qquad \chi{\cal L}\frac{\del{\cal L}}{\del\phi^i} -
  \del_a\left(\chi{\cal L}\frac{\del{\cal
  L}}{\del(\del_a\phi^i)}\right) = 0.
\eeq
Usually, this equation will reduce to an identity by use of equation
(\ref{eq:chi-motion}), ${\cal L}=0$. But it \emph{does} give non-trivial 
equations in cases where  
$\frac{\del{\cal L}}{\del\phi^i}\sim \frac{1}{{\cal L}}$
or $\frac{\del{\cal L}}{\del(\del_a\phi^i)}\sim \frac{1}{{\cal
L}}$.  (Then the factors ${\cal L}$ are eliminated from equation
\ref{eq:phi-motion}.) 
This is a special situation, however, and this approach thus has
limited applicability.

\paragraph{Generic example: The point particle}

The action of a relativistic point particle is
\beq
  \label{eq:pointparticleaction}
  S =  m\int d\tau\sqrt{ -\dot{X}^\alpha \dot{X}_\beta} 
	= m\int d\tau\sqrt{-\dot{X}^2},
\eeq
where $\dot{X}^\alpha \equiv  
\frac{dX^\alpha}{d\tau}$, and $\tau$ is some parametrization of the
world-line.

A massless limit can now be found as described above. Introduce the
auxiliary field $\chi$ and write
\begin{eqnarray}
  \nonumber
  S_\chi &=& \frac{1}{2}\int d\tau(\chi {\cal L}^2+ \frac{m^2}{\chi})
  \\ \label{eq:ppI_csaction}
  &       =& \frac{1}{2}\int d\tau(-\chi \dot{X}^\alpha\dot{X}_\alpha + 
           \frac{m^2}{\chi}).
\end{eqnarray}
This is the manifestly reparametrization invariant form of the
point-particle action with $\chi$ acting as the einbein. Putting
$m=0$, the reparametrization invariant form of the massless particle
results,
\beq
  \label{eq:pointI-lim}
  S_\chi^{m=0} = -\12 \int d\tau \chi \dot{X}^\alpha\dot{X}_\alpha.
\eeq

\subsection{ Phase space}

This second method of arriving at an action that admits taking the $T \to 0$ 
limit is designed for constrained systems. Several applications of the method 
can be found in 
\cite{marnelius:1982,karlhede:1986,isberg:1994,lindstrom:1991,lindstrom:1991b,lindstrom:1991c,hassani:1994}.

Start again with the action (\ref{eq:generalaction}). 
Derive the canonical conjugate momenta
\beq
  \pi_i = \frac{\del L}{\del \dot{\phi^i}},
\eeq
and find the total Hamiltonian as described in 
Section \ref{sec:constrained_systems},
\beq
  H = H_{naive} + \lambda^m\theta_m.
\eeq
The derivation of the total Hamiltonian $H$ involves working out
the constraint structure, which can be a cumbersome task. 

\begin{figure}
  \begin{center}
    \fbox{\epsfig{file=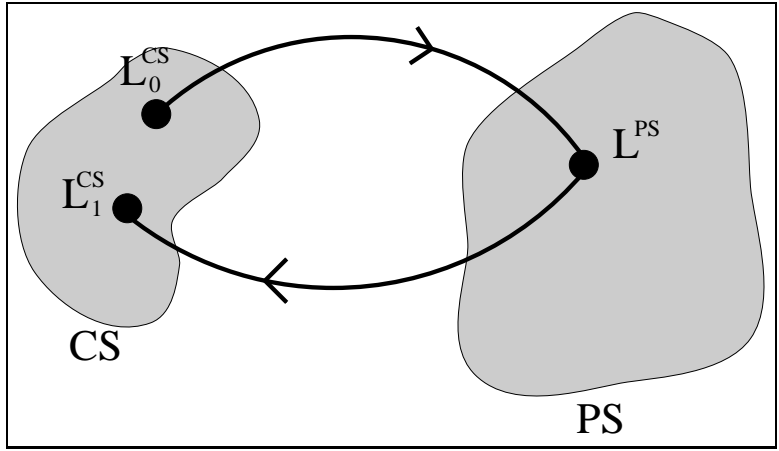, width=10cm}}
  \end{center}
  \caption{{\small Start with a Lagrangian $L^{CS}_0$ in configuration
    space (CS), and perform a Legendre transform to the phase space
    (PS) Lagrangian $L^{PS}$. Going back to configuration space
    may, for constrained systems, give a new (but equivalent) Lagrangian
    $L^{CS}_1\neq L^{CS}_0$. And although $L^{CS}_0$ is not defined in 
	the limit $T=0$, $L^ {PS}$ or $L^{CS}_1$ may be.}
  } 
  \label{fig:method2}
\end{figure}

Having found the total Hamiltonian, the phase space 
action is:
\beq
  S^{PS} = \int dx \left( \pi_i\dot{\phi^i} 
	- H(\phi, \pi, \nabla\phi) \right).
\eeq
The momenta 
$\pi_i$ can then be eliminated by solving their equations of
motion. Substituting for the solutions of $\pi_i$ gives a new 
configuration space action
\beq
  \label{eq:generalCSction}
  S^{CS} = \int dx \left[ \pi(\phi, \del \phi)\dot{\phi} - H(\phi, 
   \pi(\phi, \del \phi), \nabla \phi) \right] .
\eeq
Unless the system under study is un-constrained (giving $H=H_{naive}$)
this action will contain  new variables (the Lagrange
multipliers) compared to the original action
(\ref{eq:generalaction}).  In other words, it is different from the
original configuration space action, but still equivalent to it
(see figure \ref{fig:method2}).
In some cases (such as the point particle) this new
form makes it possible to take the
$T
\to 0$  limit, but in general the merit of this approach is that
one may take the limit at the level of constraints.

When there are no constraints
the above construction will be circular and give back the
original action. However, it may still be possible 
to make sense of the $T\to 0$ limit
in the intermediate phase space description. This is illustrated in
some of the applications in Section 3 below.

Again, the generic example which illustrates the method is
provided by the point particle.

\paragraph{The point particle}
From the action (\ref{eq:pointparticleaction}) one finds the
momenta,
\beq
  P_\alpha = \frac{\del L}{\del\dot{X}^\alpha}
  			 = -\frac{m\dot{X}_\alpha}{\sqrt{-\dot{X}^2}}.
\eeq
The naive Hamiltonian vanishes since
\beq
  H_{naive} = P_\alpha\dot{X}^\alpha - L = 0.
\eeq
This is in accordance with the diff invariance of the action 
integral (\ref{eq:pointparticleaction}) and the
discussion in section \ref{sec:diff_hamilton}

The expression for $P_\alpha$ is not invertible, leading to one primary
constraint,
\beq
  \theta = P^2 + m^2 \approx 0,
\label{masshell}
\eeq
where $P^2 \equiv P^\alpha P_\alpha$. There are no secondary constraints,
and the total Hamiltonian is
$H = \lambda(P^2 + m^2)$,
which gives the following phase space action:
\beq
  \label{eq:ppII_psaction}
  S^{PS} = \int d\tau \left(P_\alpha\dot{X}^\alpha 
  		 - \lambda(P^\alpha P_\alpha + m^2)\right).
\eeq
A variation in $P_\alpha$ gives
\beq
  \delta P_\alpha \Rarr \qquad \delta 
  		 S^{PS} = \int d\tau(\dot{X}^\alpha
		 		-2\lambda P^\alpha)\delta P_\alpha,
\eeq
which by use of Hamilton's principle leads to a solution for the momentum,
\beq
  P^\alpha = \frac{\dot{X}^\alpha}{2\lambda}.
\eeq
Substituted back into (\ref{eq:ppII_psaction}) this gives a new configuration
space action,
\beq
  S^{CS} = \12 \int d\tau\left(\frac{1}{2\lambda}
  		 \dot{X}^\alpha\dot{X}_\alpha-2m^2\lambda \right).
\eeq
A comparison with the result (\ref{eq:ppI_csaction}) found by method I
reveals that the two methods give exactly the same result, with the
identification $\chi = (-2\lambda)^{-1}$.

\section{Applications}

Both these methods have been applied to a variety of models before,
and to get an overview and further references, the reader may consult
\cite{isberg:1994}. 
In the subsequent sections of this article the methods are applied to
Weyl-invariant strings, D-strings, and to general relativity. Many
of the results are known and included as illustrations, but,
e.g., the applications to the manifestly diffeomorphism invariant
actions (\ref{eq:Pol-string}), (\ref{eq:D2-Weylaction}) and
(\ref{GR}) below, are new.

\subsection{The Bosonic string}

\paragraph{The Nambu Goto string}

The Nambu-Goto form of the action for a bosonic string is
\cite{nambu:1970,goto:1971}
\beq
  \label{eq:NG-string}
  S = T\int d^2\xi\sqrt{-\gamma},
\eeq
where $\xi^a$ parameterizes the world sheet, $\gamma \equiv
\det(\gamma_{ab})$, and $\gamma_{ab}$ is the induced metric,
$\gamma_{ab}\equiv G_{\alpha\beta}\frac{\del X^\alpha}{\del\xi^a}
\frac{\del X^\beta}{\del\xi^b}
= G_{\alpha\beta}\del_a X^\alpha\del_b X^\beta$,
$G_{\alpha\beta}$ being the spacetime metric.

In \cite{schild:1977,karlhede:1986} it is shown that in the
tensionless limit $T\to 0$ the 
Nambu-Goto string gives a spacetime conformally invariant theory
with a degenerate metric, a so-called null string. Only the results
are presented here.

The auxiliary field method gives the action
\beq
  \label{eq:bstring_m1}
  S = -\12 \int d^2\xi~\chi\gamma,
\eeq
where $\chi$ is an auxiliary field.

The phase space method  leads to a Hamiltonian which can be written
\beq
  \label{eq:NGstringHamilton}
  H = \lambda(P^2 + T^2 \acute{X}^2)+\rho P_\alpha\acute{X}^\alpha,
\eeq
where $\dot{X}^\alpha\equiv \frac{\del X^\alpha}{\del\xi^0}$, and
$\acute{X}^\alpha\equiv\frac{\del X^\alpha}{\del\xi^1}$, and
$\lambda$ and $\rho$ are Lagrange multipliers. Integrating out the
momenta from the corresponding phase space action and recombining
the Lagrange multipliers into a metric yields the manifestly
reparametrization invariant form (\ref{eq:Pol-string}) below.
Taking the limit $T\to 0$ in (\ref{eq:NGstringHamilton}) and
repeating the procedure yields
\begin{eqnarray}
  S_1 & = & \int d^2\xi~ V^aV^b\gamma_{ab},
\label{vectac}
\end{eqnarray}
where the Lagrange multipliers have been combined into the vector
density $V^a$ instead. If one chooses to also integrate out one of
the lagrange multipiliers, the result is
\begin{eqnarray}
  S_2 & = & \12 \int d^2\xi~ \frac{\gamma}{V},
\label{scalac}
\end{eqnarray}
where  $V$ is a scalar density. With the appropriate identification
this is again (\ref{eq:bstring_m1}).
Interpretations of the last two actions  are
given e.g. in
\cite{lindstrom:1991}.

\paragraph{The ``Polyakov form''}
\label{sec:weylstring}

A Weyl-invariant form of the string action that is equivalent to the
Nambu Goto action was found 
by Brink, Di Vecchia and Howe \cite{brink:1976}, 
and by Deser and  Zumino \cite{deser:1976}
It was used by Polyakov to study quantum properties of strings and 
is usually referred to as the Polyakov action. It reads
\beq
  \label{eq:Pol-string}
  S = \frac{T}{2}\int d^2\xi\sqrt{-g}g^{ab}\gamma_{ab},
\eeq
where $g_{ab}$ is an auxiliary metric, and $g\equiv\det(g_{ab})$,
and the Weyl invariance means that the action is invariant under 
rescalings $g_{ab}\to \Omega g_{ab}$ of the metric.

The auxiliary field method becomes difficult to use for this form
of the action, although by making use of the Weyl-invariance one can
indeed show that the resulting equations are equivalent to those
of (\ref{vectac}) and (\ref{scalac}). This is left as an edifying
exercise for the ambitious reader.

\paragraph{Phase space}

Before applying the phase space method to the action
(\ref{eq:Pol-string}) a few remarks are needed.

First,  when viewed in the ``center of
momentum'' frame where $P_\alpha =(E,\underline{0})$, the limit 
$m\to 0$ in (\ref{masshell}), or $T\to 0$ in
(\ref{eq:NGstringHamilton}) is equivalent to neglecting the
rest-mass or tension, respectively, compared to the energy $E$.
(This is why the limits discussed may be labelled ``high energy
limits''). Formally, then, the limits may be achieved by
neglecting terms proportional to $m$ or $T$ compared to those
proportional to the momentum $P$.

Second, in the manifestly reparametization invariant formulation of
the theory, there is no longer a clear separation into, e.g.,
$T$-dependent and $T$-independent terms. Therefore a new approach
is needed.

In view of these remarks, the strategy to be adopted is to rescale
to dimensionless fields and then ignore terms independent of $P$.
This gives an unambigous way of finding the limit of large momenta.

Rescaling $X^\alpha\to \sqrt{T}X^\alpha$, the action
(\ref{eq:Pol-string}) becomes
\beq
  \label{eq:Pol-string2}
  S = \frac{1}{2}\int d^2\xi\sqrt{-g}g^{ab}\gamma_{ab},
\eeq
The momenta are:
\begin{eqnarray}
\label{momenta}
   P_\alpha & \equiv & \frac{\del L}{\del \dot{X}^\alpha} 
   =  \sqrt{-g}(g^{00}\dot{X}_\alpha + g^{10}\acute{X}_\alpha).
\end{eqnarray}
In addition, the momenta
 $\Pi^{ab}$ for the auxiliary metric $g_{ab}$ vanishes
everywhere. This is consistent with the phase
space derivation where the metric arose as a combination of Lagrange
multipliers. It is safe to ignore its momentum constraint and
time-derivative
in the construction of the Hamiltonian. Unlike the cases
previously discussed, (\ref{momenta}) is nondegenerate and may be
solved for $\dot{X}^\alpha$:
\beq
  \label{eq:PolinverseP}
  \dot{X}_\alpha = \frac{1}{g^{00}}\left( \frac{P_\alpha}{T\sqrt{-g}}-
    g^{10}\acute{X}_\alpha \right).
\eeq
The fact that this is possible further means that the momenta are
independent functions of $\dot X^\alpha$. Thus there are no
constraints in the theory.

The Hamiltonian now follows directly. In absence
of constraints the total Hamiltonian equals the naive Hamiltonian. 
\begin{eqnarray}
  \nonumber
  H & = & P_\alpha\dot{X}^\alpha - L
  \\ \nonumber
  & = & P_\alpha \dot{X}^\alpha - \12
\sqrt{-g}(g^{00}\dot{X}^\alpha\dot{X}_\alpha 
  + 2g^{01}\dot{X}^\alpha\acute{X}_\alpha
  +g^{11}\acute{X}^\alpha\acute{X}_\alpha)
  \\ \nonumber
  & =& \frac{1}{2g^{00}\sqrt{-g}} (P^2 +  \acute{X}^2)-
  \frac{g^{01}}{g^{00}} P\cdot \acute{X}.
\end{eqnarray}
Dropping the term independent of $P$ (the term $\propto
\acute{X}^2$) and integrating out the momenta from the phase space
action yields
\beq
\label{csact}
S=2\int
d^2\xi\left[{(g^{01})^2}\over{g^{00}}\acute{X}^2+2g^{01}\dot{X}\cdot\acute{X}
+g^{00}\dot{X}^2\right].
\eeq
The action (\ref{csact}) is nolonger reparametrization invariant
with 
$g^{00}$ and $g^{01}$ transforming as components of a second rank 
tensor. However, making the field redefinitions 
\beq
V^0\equiv \sqrt{2\over{g^{00}}}(-g)^{\frac{1}{4}}g^{01}, \qquad V^1\equiv
\sqrt{2g^{00}}(-g)^{\frac{1}{4}},
\eeq
we recover the reparametrization invariant action (\ref{vectac}) 
in terms of the vector density $V^a$. This illustrates how
unconstrained 
models may be handled.

\subsection{The D-string}

D-branes are extended objects that arise naturally in
string theory. They are defined by the property that open strings can end
on them, but have their own dynamics. A D-string is another name for a
D1-brane i.e. a 2-dimensional D-brane.

The fluctuations of the D-string is described by the Dirac-Born-Infeld (DBI)
action, so called because of its similarities to the Dirac action for
membranes and to the Born-Infeld action
\cite{born:1934}. Disregarding the dilaton field, it is
\cite{polchinski:1998}
\beq
  \label{eq:D2-brane}
  S= T\int d^2\xi~
	\sqrt{-\det(\gamma_{ab}+B_{ab}+F_{ab})},
\eeq
where $T\equiv\frac{1}{2g\pi\alpha'}$ is a constant, $g$ is the string
coupling and $2\pi\alpha'$ is the inverse of the fundamental string
tension, and\footnote{ 
  Remember that $\del_a = \frac{\del}{\del\xi^a}$ and 
  $\del_{[a}A_{b]}=\del_a A_b- \del_b A_a$.
}
$F_{ab}= 2\pi\alpha'\del_{[a}A_{b]}$, $A_a$ being an abelian gauge field.
Furthermore,
\beq
  \gamma_{ab}(\xi)\equiv \del_aX^\alpha\del_bX^\beta G_{\alpha\beta}(X); \qquad
  B_{ab}(\xi) \equiv \del_aX^\alpha\del_bX^\beta {\cal B}_{\alpha\beta}(X)
\eeq
are the induced metric and antisymmetric tensor pulled back to the brane.
$G_{\alpha\beta}$ is the background (symmetric) metric, and 
${\cal B}_{\alpha\beta}$ is the background (antisymmetric) Kalb-Ramond field.
(Hitherto $G$ has been taken to be constant and ${\cal B}$ to vanish in the
discussion of strings.)

Because of the relation between $T$ and $g$ the $T\to 0$
limit may be viewed as a strong coupling limit where $g\to\infty$ and
$\alpha'$ is held fixed.

The independent fields in this action are the embedding
$X^\alpha(\xi)$ and the gauge fields $A_a(\xi)$. If $A_a = 0$ the
D-string reduces to the bosonic string.

\paragraph{The DBI action}

Defining $M_{ab}\equiv \gamma_{ab}+B_{ab}+F_{ab}$,
the tensionless limit found using an auxiliary field is easily seen to be
\beq
  \label{eq:BI-limI}
  S^{T=0}_\chi = -\12\int d^{p+1}\xi~ \chi \det(M_{ab}).
\eeq
The equation of motion for $\chi$ is found from a variation
$\delta\chi$:
\beq
  \delta \chi \Rarr \quad
	\det(M_{ab}) = 0.
\eeq 
This is similar to what was found for the bosonic string. But in
the present case the degeneracy does not imply that the world volume is
a null surface. It merely gives a relation between the
$X^\alpha$ and $A_a$ fields.

The phase space approach described in \cite{lindstrom:1997} gives a Hamiltonian
\beq
  \label{eq:BI-hamilton}
  H = P^a\del_a A_0 
	+ \lambda\Theta_A + \rho^i \Theta_i + \sigma\Theta_B 
	+ \tau\Theta_C,
\eeq
where $P^a$ is the canonical conjugate momentum to $A_a$, and
$\lambda$, $\rho$, $\sigma$, and $\tau$ are Lagrange multipliers, and
(with ${\cal B}_{\alpha\beta}$ set to zero)
\begin{eqnarray}
  \Theta_1 &\equiv& \Pi_\alpha\del_1 X^\alpha 
	+ \frac{P^1}{2\pi\alpha'} F_{11} \approx 0,
  \\
  \Theta_A &\equiv& \Pi_\alpha\Pi^\alpha 
	+ \frac{P^1\gamma_{11}P^1}{(2\pi\alpha')^2} + T^2\det(M_{11})
	\approx 0,
  \\
  \Theta_B &\equiv& P^0 \approx 0,
  \\
  \Theta_C &\equiv& \del_c P^c \approx 0,
\end{eqnarray}
are the
constraints, with $\Pi_\alpha$ the conjuagate momentum to
$X^\alpha$.

Integrating out the momenta from the phase space action and 
redefining fields in terms of the Lagrange multipliers gives a 
manifestly diffeomorphism invariant form of (\ref{eq:D2-brane}), 
just as for the ordinary string. The resulting action is
given in (\ref{eq:D2-Weylaction}) below.

Taking the $T\to 0$ limit in (\ref{eq:BI-hamilton})  and repeating 
the procedure results in  the tensionless limit of the DBI-string:
\beq
\label{eq:DBI-limit2}
  S_2^{T=0} = \frac{1}{4} \int d^{p+1}\xi V^aW^bM_{ab},
\eeq
where $V^a$ and $W^a$ are the new vector density fields defined from
the Lagrange multipliers. Integrating out one of the Lagrange  multipliers in
addition to the momenta yields
\begin{eqnarray}
  \label{eq:DBI-limit1}
  S_1^{T=0} &=& \frac{1}{4} \int d^{p+1}\xi V\det(M_{ab}),
\end{eqnarray}
where $V$ is a scalar density.   The result would essentially be the 
same with the background field ${\cal B}_{\alpha\beta}$ included. 
The second action (\ref{eq:DBI-limit1}) is identical to what was found
in (\ref{eq:BI-limI}). The action (\ref{eq:DBI-limit2}) can be shown
\cite{lindstrom:1997} to  imply that the world
surface of the D-string generally splits into a collection of tensile
strings or, in special cases, massless particles. Thus it leads to a
parton picture of D-branes in this limit.

\paragraph{The Weyl-invariant form}

An equivalent, but Weyl-invariant form of the D-string action was 
constructed in \cite{lindstrom:1988} 
\beq
  \label{eq:D2-Weylaction}
  S = \frac{T}{2}\int d^2\xi\sqrt{-s}s^{ab}M_{ab},
\eeq
where $s\equiv\det(s_{ab})$ and $s_{ab}$ is an auxiliary tensor field
with no symmetry assumed. Elimination of $s_{ab}$ gives the
DBI action (\ref{eq:D2-brane}).

\paragraph{Phase space} The $T\to 0$
limit of the action (\ref{eq:D2-Weylaction}) can only be meaningfully 
discussed in the phase space approach. As discussed immediately
before equation  (\ref{eq:Pol-string2}), the starting point should
be the action  (\ref{eq:D2-Weylaction}) with the fields rescaled to
be dimensionless. In this case a convenient rescaling is
$X^\alpha\to
\sqrt{T}X^\alpha$ and $A_a\to 2\pi \alpha 'TA_a$. Instead of doing this, 
the discussion below shows that the Hamiltonian can be brought to the 
same form as that of the DBI action, equation (\ref{eq:BI-hamilton}). 
The subsequent limit $T\to 0$ will then be the same.
The fields to be considered as independent variables in the
action (\ref{eq:D2-Weylaction}) are $X^\alpha$, $A_a$ and $s_{ab}$. 
Setting ${\cal B}_{\alpha\beta}=0$, the canonical conjugate momenta 
for these fields are.
\begin{eqnarray}
  \Pi_\alpha & \equiv & \frac{\del L}{\del \dot{X}^\alpha}
   =  T \sqrt{-s} \left( s^{00} \dot{X}_\alpha + \12(s^{01}+s^{10})X_\alpha ' 
        \right)
  \label{eq:BImomentumPi}
  \\
  P^a & \equiv & \frac{\del L}{\del \dot{A}_a} 
   =  T \sqrt{-s} \12(s^{0a} - s^{a0})2\pi\alpha '
  \label{eq:BImomentumP}
  \\
  \Sigma^{ab} & \equiv & \frac{\del L}{\del \dot{s}_{ab}} = 0
  \label{eq:BImomentumSigma}
\end{eqnarray}
The first equation (\ref{eq:BImomentumPi}) is invertible which means that 
an explicit expression for $\dot{X}^\alpha$ can be obtained:
\beq
  \dot{X}_\alpha = \frac{1}{s^{00}} \left(\frac{\Pi_\alpha}{T\sqrt{-s}}
        -\12(s^{01}+s^{10})X_\alpha ' \right) 
\eeq
The second equation (\ref{eq:BImomentumP}) is obviously not
invertible since the momentum $P$  is completely
independent of the fields $A$.
Its definition results in the following  constraints
\begin{eqnarray}
  \Theta_0 &\equiv& P^0 \approx 0,
  \\
  \Theta_1 &\equiv& P^1 -
	  \frac{T}{2}\sqrt{-s}(s^{01}-s^{10})2\pi\alpha' \approx 0.
\end{eqnarray}
The last equation (\ref{eq:BImomentumSigma}) says that the conjugate
momenta to $s_{ab}$ are
identically zero. This reflects the fact that 
$s_{ab}$ are non-dynamical variables to be treated on the same
footing as Lagrange multipliers (cf. the comment following
(\ref{momenta})).

Now,  derive the naive Hamiltonian. Disregarding 
$\Sigma^{ab}$,
\begin{eqnarray}
  \nonumber
  H_{naive} & = &
        \Pi_\alpha\dot{X}^\alpha + P^a\dot{A}_a - 
        \frac{T}{2}\sqrt{-s}s^{ab}(\gamma_{ab} 
	+ 2\pi\alpha'(\del_aA_b - \del_bA_a)).
\end{eqnarray}
Some rearrangements yield a simpler form,
\begin{eqnarray}
  \nonumber     
  H_{naive} &=& \frac{1}{2Ts^{00}\sqrt{-s}}\left( \Pi_\alpha\Pi^\alpha 
        + \frac{P^1\gamma_{11}P^1}{(2\pi\alpha')^2} 
	+ T^2\gamma_{11}\right) 
  \\    
  & &   - \frac{s^{01}+s^{10}}{2s^{00}}\Pi_\alpha X'^\alpha
        + P^a\del_a A_0.
\end{eqnarray}
The consistency condition on the primary constraint $\Theta_0$ gives a
secondary ``Gauss law'' constraint 
\beq
  \Theta_2\equiv \del_aP^a\approx 0,
\eeq
while $\Theta_1$ gives nothing new.
There are no tertiary constraints.
The four component fields of $s_{ab}$ are Lagrange multipliers which
can be redefined as
\begin{eqnarray}
  \lambda & \equiv & \frac{1}{2Ts^{00}\sqrt{-s}},
  \\
  \rho & \equiv & -\frac{s^{01}+ s^{10}}{2s^{00}},
  \\
  \varphi & \equiv & \frac{T}{2}\sqrt{-s}(s^{01}-s^{10}).
\end{eqnarray}
Including the constraints, the total Hamiltonian is obtained and reads
\begin{eqnarray}
  \nonumber
  H &=& 
	\lambda (\Pi^2 + \frac{P^1\gamma_{11}P^1}{(2\pi\alpha')^2}
	+T^2\gamma_{11})
	+\rho \Pi_\alpha\del_1X^\alpha
	+ P^a\del_aA_0
	\\&&	
	+\sigma_0P^0
	+\sigma_1(P^1+\varphi)
	+\tau\del_aP^a
\end{eqnarray}
The phase space Lagrangian is 
$L^{PS}=\Pi_\alpha\del_0X^\alpha+P^a\del_0A_a-H$, and a variation of $\varphi$
gives $\sigma_1=0$, which means that the constraint
$\Theta_1=P^1+\varphi$
in fact makes no difference.
Then it is obvious that this is exactly the same Hamiltonian and phase space
Lagrangian as was found
from the DBI action for the D-string
(\ref{eq:BI-hamilton}).

 A rescaled version of the above expressions
is achieved by putting $T=1$, and the limit $T\to 0$ is then equivalent to dropping
terms that do not contain the momenta. The exact procedure follows from the
discussion in \cite{lindstrom:1997} and results in the actions
(\ref{eq:DBI-limit2}) and (\ref{eq:DBI-limit1}), as mentioned above.

The treatment of the Weyl-invariant $D$-string parallels that of the
Polyakov string versus the Nambu-Goto string.
This is no surprise, since the string
action can be seen as a special case of the D-string action where
$A_a= {\cal B}_{\alpha\beta}= 0$, and $s_{ab}$ is symmetric.

\subsection{General relativity}

The Hilbert action describing the dynamics of spacetime is
\beq
  S[g] = \frac{1}{\kappa}\int d^4x\sqrt{-g} g^{\alpha\beta}
R_{\alpha\beta},
\label{GR}
\eeq
where $\kappa$ is a constant, $g\equiv\det(g_{\alpha\beta})$, 
$g_{\alpha\beta}$ is the metric, and
$R_{\alpha\beta}$ is the Ricci tensor (a function of the metric and
its first and second derivatives). 
The limit to be studied in this case is $\kappa\to\infty$.
Einstein's constant $\kappa$ is defined as 
$\kappa \equiv \frac{8\pi}{c^3}G_N$, so this limit can be thought of
as either a limit where Newton's gravitational constant $G_N\to\infty$ or
a limit where the speed of light $c\to 0$.
As the speed of light approaches zero, lightcones will collapse into
spacetime lines, and points in space become disconnected. So this
limit leads to an ultralocal field theory.

Yet another interpretation of this limit comes from the
observation (which is possible in the Hamilton formulation) that it is
equal to the \emph{zero signature limit}, which represents some
intermediate stage between Euclidean space (signature $+1$) and Minkowski
space (signature $-1$) \cite{henneaux:1979}.

\air
Although the action above in appearance very much resembles the Polyakov
string action (\ref{eq:Pol-string}), it is considerably more 
intricate to analyze. Nevertheless the limit can be found using the phase space
approach in suitable coordinates.

\paragraph{Phase space} 

A fruitful approach to a Hamiltonian description of general
relativity is the ADM approach \cite{arnowitt:1982}. The crucial step
is the introduction of a new set of variables, replacing the 10
independent components of the metric $g_{\alpha\beta}$.
The new variables are the \emph{lapse function} $N$, the \emph{shift
functions} $N_{\mu}$, and the 3-dimensional induced metric (on a
hypersurface $\Sigma$) $h_{\mu\nu}$, with the following relation to the four
dimensional metric:
\beq
  \label{eq:new-vars}
  g_{\alpha\beta} = \left(
  \begin{array}{cc}
	N_\mu N^\mu-N^2	& N_\nu \\
	N_\mu^T		& h_{\mu\nu}
  \end{array} 
  \right);\qquad 
  \begin{array}{l}
	\alpha,\beta=0,1,2,3;\\
	\mu,\nu=1,2,3.
  \end{array}
\eeq
Through some reformulations, and the derivations of the canonical
conjugate momenta, the Hamiltonian is found to be
\beq
   \label{eq:GRhamilton}
   H = \tilde{N}\theta + N_\nu \theta^\nu,
\eeq
where $\tilde{N}\equiv\kappa N$, 
and $\theta$ and $\theta^\nu$ are to be considered as constraints,
\begin{eqnarray}
  \theta & = & \frac{1}{\sqrt{h}}(\pi^{\mu\nu}\pi_{\mu\nu} - \12\pi^2) 
           - \frac{1}{\kappa^2} \sqrt{h} R_\Sigma,
  \\
  \theta^\nu & = & -2D_\mu\pi^{\mu\nu}.
\end{eqnarray}
Here, $\pi^{\mu\nu}$ is the canonical conjugate momentum to $h_{\mu\nu}$,
$R_\Sigma$ is the Ricci scalar on the hypersurface $\Sigma$ and
$D_\mu$ is the covariant derivative on $\Sigma$.

Taking the limit $\kappa\to\infty$ is now possible by just dropping
the last term in $\theta$. (Note that if we rescale to dimensionless 
coordinates in (\ref{GR}) this again amounts to dropping terms not 
containing momenta.)
And as mentioned above, this has the same effect as taking
the zero signature limit $\varepsilon \to 0$. The signature
$\varepsilon$ of the spacetime metric only
influences this term, and enters in such a way that taking
$\varepsilon \to 0$ removes the term proportional to $\sqrt{h}R$
\cite{teitelboim:1980}.

This limit has in turn been shown \cite{henneaux:1979} to correspond
to the four-dimensional action
\beq
  \label{eq:e0GRaction}
  S= \int d^4x \Omega(x) ({\cal K}^{\alpha\beta}{\cal K}_{\alpha\beta}
     - {\cal K}^2); \quad \alpha,\beta=0,1,2,3.
\eeq
The equivalence can be demonstrated by showing that this action
gives the same Hamilton formulation (\ref{eq:GRhamilton}) as the
$\varepsilon\to 0$ limit of the general relativity action.

The independent fields in the action (\ref{eq:e0GRaction}) are the
positive scalar density $\Omega(x)$ and the components of a symmetric
covariant tensor $\tilde{g}_{\alpha\beta}(x)$.
This ``metric'' $\tilde{g}_{\alpha\beta}$ is degenerate, i.e. 
$\det(\tilde{g}_{\alpha\beta})=0$, which means that it has at most 9
independent components. Together with $\Omega$ this gives 10
independent fields, which is the same number as in the original
action.

${\cal K}_{\alpha\beta}$ is the second fundamental tensor, defined as the 
Lie derivative in a unique direction $\vec{e}$,
\beq
{\cal K}_{\alpha\beta} = \12 \pounds_e \tilde{g}_{\alpha\beta}
  =\12 (e^\gamma \tilde{g}_{\alpha\beta,\gamma} 
         +e^\gamma_{,\alpha} \tilde{g}_{\gamma\beta}
         +e^\gamma_{,\beta} \tilde{g}_{\alpha\gamma}).
\eeq
And the vector field $\vec{e}$ is defined through
\beq
  {\cal G}^{\alpha\beta} = \Omega^2e^\alpha e^\beta; \qquad
  {\cal G}^{\alpha\beta} \equiv 
	\frac{1}{3!}\varepsilon^{\alpha\gamma\delta\epsilon}
	\varepsilon^{\beta\zeta\eta\theta}
\tilde{g}_{\gamma\zeta}\tilde{g}_{\delta\eta}\tilde{g}_{\epsilon\theta}.
\eeq
Hence ${\cal G}^{\alpha\beta}$ is  the minor of
$\tilde{g}_{\alpha\beta}$. The vector $\vec{e}$ is completely
determined from
$\tilde{g}_{\alpha\beta}$ and $\Omega$. 
It satisfies $\tilde{g}_{\alpha\beta}e^\beta=0$ and is thus orthogonal
to any other vector $v^\alpha$, i.e.  
$\tilde{g}_{\alpha\beta}e^\alpha v^\beta = 0$.

Since the metric is degenerate (i.e. the determinant vanishes), there
is no inverse $\tilde{g}^{\alpha\beta}$ satisfying
$\tilde{g}^{\alpha\beta}\tilde{g}_{\beta\gamma}=\delta^\alpha_\gamma$. 
However, the class of symmetric tensors $G^{\alpha\beta}$ defined by
\beq
  G^{\alpha\beta}\tilde{g}_{\beta\gamma} = \delta^\alpha_\gamma -
	\Theta_\gamma e^\alpha,
\eeq
where $\Theta_\gamma$ is an arbitrary vector satisfying 
$\Theta_\gamma e^\gamma=1$, may instead be used to raise indices which
makes ${\cal K} = G^{\alpha\beta}{\cal K}_{\alpha\beta}$ and
${\cal K}^{\alpha\beta}{\cal K}_{\alpha\beta}$
well defined. 

For an elaboration of the ideas presented above, the reader should
consult \cite{henneaux:1979}.

\air
The result of this section is that it \emph{is} possible to find an
interesting limit using the phase space method. This is not as
straight forward as in the string cases, but by introduction of
the more suitable ADM coordinates, it can be done.

The most troublesome part is the step from phase space
to configuration space. This can not be  easily done by integrating out
momenta and redefining the Lagrange multipliers as before. Instead, the idea is
to make an ansatz for a configuration space action, and then to
show that it gives the right Hamiltonian. 

As for the string models, it was found that the limit corresponds to a
degenerate geometry. In the present case, this means a non-Riemannian
space halfway between Euclidean and Minkowski space,
which corresponds to a theory of gravity based, not on local
Poincar\'e invariance, but on local \emph{Caroll invariance}.

\section{Conclusions}

One aim of this article was to present  methods for deriving
tensionless limits of strings and the analogue in other models. This
has been done by giving a general description as well as several
explicit calculations.

An interesting result that was given in the introduction is the derivation
of the 
naive Hamiltonian for diffeomorphism invariant theories with only tensor 
fields. It was demonstrated that for such models the Hamiltonian is
constrained to be zero when the fields satisfy the field equations.

A second goal of the article has been to investigate the applicability
of the methods.
It was known from before that they work well for a number of
string models. The derivations here has revealed that the methods work 
perfectly also if
we start from the Weyl-invariant form of the bosonic string and
D-string actions. These are models that already at the very beginning
contain Lagrange multipliers (an auxiliary metric). 
The methods apply in such cases as well, which  emphasizes their generality. 

Applying the methods to General Relativity turned out not to be
quite  straight forward. However, by changing to ADM coordinates, 
the phase space treatment yielded expressions that could be handled.

A general remark is that in the tensionless limit, the string
models provide conformally invariant theories, and that the
geometries turn out to be degenerate. That the point particle action 
has spacetime conformal symmetry is well known. This result is 
directly generalizable to the other brane-actions discussed and is 
explicitly shown in the cited literature on
the $T\to 0$ limit. The same holds for the degeneracy. A massless 
point particle moves on a null geodesic, a tensionless string moves on 
a null-surface. The induced metric is thus degenerate as is the metric 
in the $\kappa\to \infty$ limit of General relativity.

\paragraph{Acknowledgement}  We thank Bj\"orn Brinne and Rikard
von Unge for comments. The work of U L was supported in part
by a grant from the Swedish Natural Sciences Research Council, and
by ``The European Superstring Theory Network'', part of the EU
program for Training and Mobility of Researchers (TMR). H G S
gratefully acknowleges support from the Nordic Council of Ministers
in the form of a Nordplus grant.

\paragraph{Notation}
The spacetime dimension is denoted $D$, and the world-volume dimension
is denoted $d$. (For particles $d=1$, and for strings $d=2$.)
To refer to components of world-volume variables, small Latin indices
are used: From the beginning of the alphabet for general
components, and middle alphabet letters when only spatial
components are referred to, i.e.
$a,b = 0,\dots, d-1$; $m,n = 1,\dots, d-1$.

Spacetime variables are denoted by Greek indices in the same
way,$\alpha,\beta = 0, \dots, D-1$; $\mu,\nu = 1,\dots, D-1$.


\bibliographystyle{unsrt}
\bibliography{physics.bib}


\end{document}